\documentclass[a4paper,12pt]{article}
\pdfoutput=1

\usepackage{cite}
\usepackage{amsmath}
\usepackage{amsfonts}
\usepackage{amssymb}
\usepackage{graphicx, rotating}
\usepackage{epsfig}
\usepackage{latexsym}
\usepackage{graphicx}
\usepackage{color}
\usepackage{amsmath,bm,amssymb}

\usepackage{color}
\usepackage[english]{babel}
\usepackage{hyperref}

\newcommand{\Slash}[1]{{\ooalign{\hfil#1\hfil\crcr\raise.167ex\hbox{/}}}}

\newcommand{\beq}{\begin{equation}}  \newcommand{\eeq}{\end{equation}}
\newcommand{\bef}{\begin{figure}}  \newcommand{\eef}{\end{figure}}
\newcommand{\bec}{\begin{center}}  \newcommand{\eec}{\end{center}}
  
\newcommand{\laq}[1]{\label{eq:#1}}  

\newcommand{\Eq}[1]{Eq.~(\ref{eq:#1})}
\newcommand{\Eqs}[1]{Eqs.~(\ref{eq:#1})}
\newcommand{\eq}[1]{(\ref{eq:#1})}

\newcommand{\ab}[1]{\left|{#1}\right|}
\newcommand{\vev}[1]{ \left\langle {#1} \right\rangle }

\def\({\left(}
\def\){\right)}

\def\O{\mathcal{O}}

\newcommand{\AND}{~{\rm and}~}
\newcommand{\EV}{ {\rm \, eV} }
\newcommand{\KEV}{ {\rm \, keV} }
\newcommand{\MEV}{ {\rm \, MeV} }
\newcommand{\GEV}{ {\rm \, GeV} }
\newcommand{\TEV}{ {\rm \, TeV} }

\def\a{\alpha}
\def\b{\beta}

\def\d{\delta}
\def\e{\epsilon}
\def\f{\phi}
\def\g{\gamma}

\def\m{\mu}

\def\D{\Delta}
\def\G{\Gamma}

\def\L{\Lambda}

\def\P{\Psi}

\def\ol{\overline}
\def\tl{\tilde}
\def\*{\dagger}

\begin{document}

\begin{center}

\vspace{1.5cm}

{\Large\bf  Opening the 1\,Hz axion window}
\vspace{1.5cm}

{\bf  David J. E. Marsh$^{1}$ and Wen Yin$^{2,3}$}

\vspace{12pt}
\vspace{1.5cm}
{\em 
$^{1}$Institut fur Astrophysik, Georg-August-Universit\"{a}t, \\ Friedrich-Hund-Platz 1, D-37077 G\"{o}ttingen, Germany \\
$^{2}$Department of Physics, Faculty of Science, The University of Tokyo,\\ 
{ Bunkyo-ku, Tokyo 113-0033, Japan\\
$^{3}$}Department of Physics, KAIST, Daejeon 34141, Korea \vspace{5pt}}

\vspace{1.5cm}
\abstract{
An axion-like particle (ALP) with mass $m_\f \sim 10^{-15}\text{ eV}$ oscillates with frequency $\sim$1 Hz. This mass scale lies in an open window of astrophysical constraints, and appears naturally as a consequence of grand unification (GUT) in string/M-theory. However, with a GUT-scale decay constant such an ALP overcloses the Universe, and cannot solve the strong CP problem. In this paper, we present a two axion model in which the 1 Hz ALP constitutes the entirety of the dark matter (DM) while the QCD axion solves the strong CP problem but contributes negligibly to the DM relic density. The mechanism to achieve the correct relic densities relies on low-scale inflation ($m_\f\lesssim H_{\rm inf}\lesssim 1 \text{ MeV}$), and we present explicit realisations of such a model. The scale in the axion potential leading to the 1 Hz axion generates a value for the strong CP phase which oscillates around $\bar{\theta}_{\rm QCD}\sim 10^{-12}$, within reach of the proton storage ring electric dipole moment experiment. The 1 Hz axion is also in reach of near future laboratory and astrophysical searches.
}

\end{center}
\clearpage

\setcounter{page}{1}
\setcounter{footnote}{0}

\section{Introduction}

Axions are hypothetical particles predicted both from theoretical and phenomenological points of view. 
 String theory and M-theory compactifications predict a so-called ``axiverse''~\cite{Witten:1984dg, Svrcek:2006yi,Conlon:2006tq,Arvanitaki:2009fg,Acharya:2010zx, Higaki:2011me, Cicoli:2012sz,Demirtas:2018akl}: low-energy effective theories with large numbers of axion-like particles (ALPs, or simply ``axions'') with a wide range of possible masses. In certain compactifications, the typical decay constants for closed string axions are around the GUT scale $f_\f 
\sim 10^{16}\GEV$.  
There, a QCD 
axion~\cite{Peccei:1977hh,Peccei:1977ur,Weinberg:1977ma,Wilczek:1977pj}, which solves 
the strong CP problem, may naturally exist as one of the string axions (see Refs.~\cite{Kim:2008hd,Wantz:2009it,Ringwald:2012hr,Kawasaki:2013ae,Marsh:2015xka} for reviews). 
The axion(s), if light enough, is long-lived and can play the role of dark matter. 

The existence of certain light axions can be excluded due to the phenomenon of ``black hole superradiance'' (BHSR)~\cite{Arvanitaki:2010sy, Arvanitaki:2014wva, Cardoso:2018tly, Stott:2018opm}. If the axion Compton wavelength is of order the BH horizon size, then spin is extracted from the BH by the Penrose process. Thus the measurement of non-zero spins (e.g. from binary inspiral gravitational waveforms or accretion disk X-ray spectroscopy) for stellar and supermassive BHs (SMBHs) set quite general exclusion limits on the axion mass, $m_\f$, as 
$
 10^{-19}\EV \lesssim m_\f\lesssim10^{-16}\EV, \AND 6\times 10^{-13}\EV \lesssim m_\f\lesssim 2\times 10^{-11}\EV,
$
respectively~\cite{Arvanitaki:2014wva,Stott:2018opm} (assuming the axion quartic self-coupling is negligible, which is a good assumption for $f_\f\sim 10^{16}\text{ GeV}$). However, the non-observation of any intermediate mass BHs ($M\sim 10^4 M_\odot$), never mind the measurement of their spins, allows for the existence of the \emph{1\, Hz axion window}:
\beq
10^{-16}\EV \lesssim m_\f\lesssim 6\times 10^ {-13}\EV \Rightarrow 0.15\text{ Hz}\lesssim \nu_\f\lesssim 910 \text{ Hz}.
\eeq
It is, on the other hand,  known that the existence of an axion may lead to the accelerated growth of BHs~\cite{Arvanitaki:2010sy}. Thus if there are many axions within the 1Hz window, not only the null observation of (spinning) IMBHs but also  the observation of SMBHs existing at red-shift $z\gtrsim 6$~\cite{Shapiro:2004ud} may be explained (see also Ref.~\cite{Volonteri:2010wz}).  The 1 Hz window is also the target of several direct detection searches~\cite{Kahn:2016aff,Ouellet:2018beu,Ouellet:2019tlz,Graham:2013gfa,JacksonKimball:2017elr,Garcon:2019inh,Nagano:2019rbw, Michimura:2019qxr}.

From the point of view of model building, the 1 Hz axion cannot be the QCD axion~\cite{Peccei:1977hh,Peccei:1977ur,Weinberg:1977ma,Wilczek:1977pj}, if the scale of spontaneous symmetry breaking is below the Planck scale, and so the 1 Hz axion alone does not solve the strong CP problem. In a M-theory compacitification, the axion potential is $V(\phi)=\Lambda^4 U(\phi/f_\f)$ and the scale
\beq 
\Lambda^4 \sim \exp[{K}/2] m_{3/2}M_{pl}^3e^{-2\pi V_\f}\laq{Lambda}
\eeq 
is generated from membrane instantons up to an $\O(1)$ coefficient, where $m_{3/2}= \O(1-100)\text{ TeV}$ is the gravitino mass,\footnote{We take $m_{3/2}$ to be the predicted SUSY scale \cite{Okada:1990vk,Ellis:1990nz,Okada:1990gg,Haber:1990aw,Ellis:1991zd} from the current measurement of the top and Higgs masses. We omit the possibility that  $\tan\b$ is quite large or small, e.g. Ref.\,\cite{Vega:2015fna}.}   $M_{pl}$ is the reduced Planck mass, $M_{pl}=1/\sqrt{8\pi G_N}\approx 2.4\times 10^{18}\GEV$, the K\"{a}hler potential $K= -3 \log[{\cal V}]+\cdots$ with ${\cal V}\sim 5000$ the total volume, and $V_\f$ is some cycle volume. The MSSM GUT coupling $\alpha_{\rm GUT}\approx 1/27-1/25$ (logarithmically dependent on the SUSY scale) is linked to the cycle volume in string units $V_\f\approx 1/\alpha_{\rm GUT}$. The resulting dynamical scale is
is \beq
\laq{Lvis}
 \Lambda_{\rm vis}\sim (0.03-0.1)\MEV.
\eeq
Since $m_\f = \L_{\rm vis}^2/f_\f$, this immediately leads to 
\beq
m_{\f}\sim (10^{-16}-10^{-15})\EV\( \frac{10^{16}\GEV}{ f_{\phi}}\)
\eeq
Thus the M-theory axiverse predicts the 1\,Hz axion~\cite{Acharya:2010zx}.\footnote{Note that the mass prediction essentially relies on the MSSM with a soft scale $\TEV-\,$PeV which leads to $1/\a_{\rm GUT}\sim 25-27.$ The MSSM suffers from the notorious Polonyi/moduli and gravitino problems.} It was shown by Matthew J. Stott and one of the authors (DJEM)\,\cite{Stott:2018opm} that taking into account the BHSR constraints  the axion mass distribution around $1\,$Hz can be either very wide or very narrow in log-space (see Ref.\,\cite{Stott:2017hvl} for the prior distributions). When the distribution is narrow, we should have several axions with masses 
around $1\,$Hz. 

This $1\,$Hz axion window, however, is closed once we take account of the cosmological 
abundance of the axion, which is too large if $f_\f \gtrsim 10^{15}\GEV$ and an $\O(1)$ initial misalignment angle, $\theta_i$, and cannot be diluted by late time entropy production~\cite{Dine:1982ah,Steinhardt:1983ia,Lazarides:1990xp,Kawasaki:1995vt,Kawasaki:2004rx}.\footnote{The onset of the oscillation occurs when $3H(T_{\rm osc})\approx m_\f$, and the Friedmann equation $3H^2M_{pl}^2=\pi^2 g_\star T^4/30\Rightarrow T_{\rm osc}\lesssim \O(1) \text{ MeV}$, with $g_{\star}$ being the relativistic degrees of freedom (which we take from Ref.~\cite{Husdal:2016haj}).  This value of $T_{\rm osc}$ is close to big-bang nucleosynthesis (BBN), whose successful predictions would be spoiled by entropy production.} The QCD axion also has an overproduction problem if $f_\f \gtrsim 10^{12}\GEV$ and $\theta_i\sim 1$. 

In this paper, we open the $1\,$Hz window in two senses: we solve the overproduction problems and show various phenomenological perspectives. In Section~\ref{sec:1Hz_model} we show that one of the would-be $1$Hz axions with string-induced dynamical scale $\L_{\rm vis}$, can be the QCD axion, i.e. it additionally couples to the gluon colour anomaly and gets a QCD-induced potential lifting the mass to the canonical value $m_a\approx 6\times 10^{-6}(10^{12}\text{ GeV}/f_a)\text{ eV}$ (and an induced strong CP phase, which we compute). Thus, we are required to introduce a second 1\,Hz axion which does not receive additional mass from QCD, and we compute its couplings to the standard model. Both axions are found to overclose the Universe for GUT scale decay constants. In Section~\ref{sec:inflation_model} we build an explicit model of low-scale inflation which overcomes the axion over abundance problems. The QCD axion has its relic abundance diluted to nearly zero, while the second axion in the 1\,Hz window obtains the correct relic abundance. We conclude in Section~\ref{sec:discussion}.

\section{The 1\,Hz axion and the QCD Axion}
\label{sec:1Hz_model}

We consider the model with two would-be 1 Hz axions, $\f$ \AND $a$, each obtaining a potential of order $\Lambda_{\rm vis}$, one of which will be later identified as the QCD axion.\footnote{See also two-axion models with QCD axion and fuzzy dark matter~\cite{Kim:2015yna}, 
 QCD axion and heavier axion dark matter~\cite{Higaki:2014qua}, and QCD axion and ALP axion dark matter with level-crossing~\cite{Ho:2018qur}. In our scenario, the level-crossing effect is not important because the oscillation of the $1$\,Hz axion starts much after the QCD phase transition.}
As we will discuss in the last section, the inclusion of more than two axions in the 1\,Hz axion window does not change our conclusion. 

\subsection{The QCD Axion and Nucleon EDMs }
In the effective field theory (after spontaneous symmetry breaking and/or compactification) the two axion potential is:
\beq
V_{\rm 1Hz}\approx  -\Lambda_1^4\cos{[c_0\frac{\phi}{f_\phi} +c_1 \frac{a}{f_a}]}- \Lambda_2^4\cos{[c_2\frac{\phi}
{f_\phi} +c_3 \frac{a}{f_a}]}+\cdots,
\eeq
 where $\cdots$ represents the constant terms cancelling the cosmological constant, $\L_i$ are the 
dynamical scales of the high-scale non-perturbative effects discussed above leading to the 1 Hz axion mass scale, and $c_i$ real numbers. In general in M-theory, both $\L_i$ are non-vanishing~\cite{Acharya:2010zx}. 
Notice that we can eliminate the CP-phases in the cosine terms from the field redefinitions,  $\f\to \f +\a_1, a\to a+\a_2$ with certain real numbers $\a_{1,2}$, unless the two axions are aligned. In the following, we omit the case of alignment of axions as well as cancellations among the parameters, unless otherwise stated. 
In a compactified, supersymmetric theory, the coupling to gluons arises generically from the gauge kinetic function.
 Since we have the freedom to redefine the field as $a\to \cos{(\Theta)}  a + \sin(\Theta) \f, \f \to  -\sin (\Theta) a + \cos(\Theta) \f,$ with $\Theta$ being real phases, the gluon-axion coupling can be written as 
\beq
\laq{intG}
{\cal L}_{\rm int}\supset -\frac{\a_s }{8\pi } \(c_s \frac{a}{f_a}+\theta \) G\tl{G}. 
\eeq
 where $\a_s$ is the strong coupling constant, $G$ and $\tilde{G}$ are the gluon field strength and dual field strength, $c_s$ is the anomaly coefficient to the gluons, and $\theta=\O(1)$ is the CP phase coming from the standard model sector in the basis where the quark masses are real, $\arg \det[m_u m_d]=0.$ Due to the QCD phase transition $a$ acquires a potential  
\beq V_{\rm QCD}(a)\approx -\chi\cos[c_s\frac{a}{f_a}+\theta]+\cdots\eeq where $\chi\approx \(0.0756\GEV\)^4$ is the topological susceptibility~\cite{Borsanyi:2016ksw}. 
Thus the total potential is\footnote{This potential is specific to the M-theory inspiration for our model. In Type-IIB string theory, the QCD axion is more naturally realised as an open string axion, and the resulting potential does not mix the fields in the same manner.}
\beq
\laq{pot}
V=V_{\rm 1Hz}(\f,a)+V_{\rm QCD}(a).
\eeq
Without loss of generality, we take $c_s=c_0=1$ to redefine $f_a$, $f_\f$ and $c_i$ in the basis of \eq{intG}. 

\begin{figure}[!t]
\begin{center}  
     \includegraphics[width=145mm]{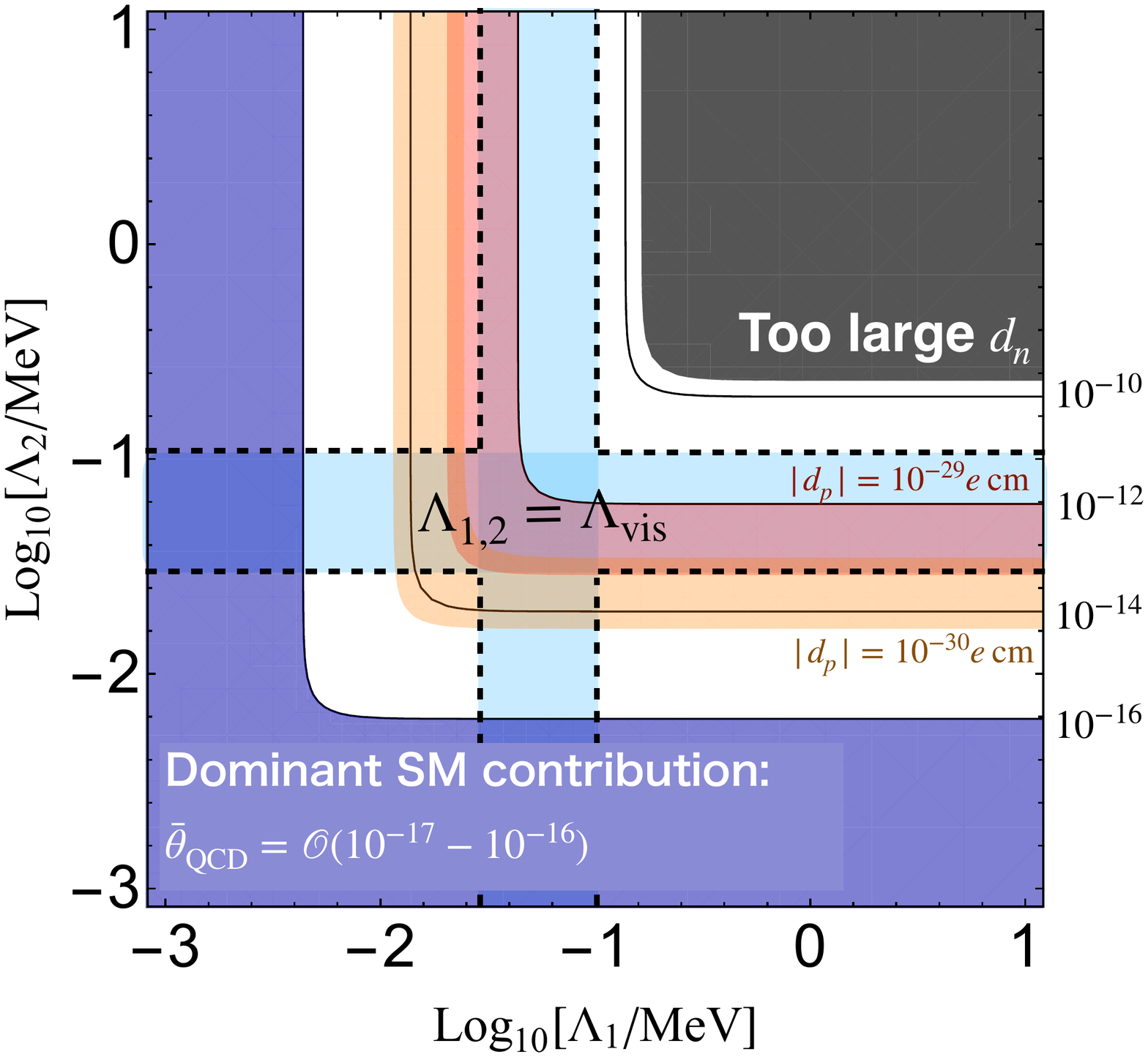}
      \end{center}
\caption{Contour plot of strong CP phase, $\bar{\theta}_{\rm QCD}$, in the two axion models varying the non-perturbative scales $\L_1,\AND \L_2$. We have fixed the other parameters as $\theta=c_1=1,c_2=0.5, c_3=2$. 
The light blue shaded region in the dashed line represents $\L_1=\L_{\rm vis}$ or/and $\L_2=\L_{\rm vis}.$  }
 \label{fig:1} 
\end{figure}

There is a general and physical CP phase $\theta$ that cannot be shifted away. The strong CP phase is related with $\theta$ as 
\beq
\bar{\theta}_{\rm QCD}= \vev{\frac{a}{f_a}}+\theta,
\eeq
where $\vev{}$ denotes the vacuum expectation value. 
By stabilizing the potential (with a quadratic approximation), at leading order in $\L_1^4, \L_2^4$ we find 
\beq
\laq{thetaCP}
\bar{\theta}_{\rm QCD}\approx \frac{\Lambda_1^4 \Lambda_2^4  (c_3-c_1c_2)^2}{\chi  \left(c_2^2 \Lambda_2^4+\Lambda_1^4\right)}\theta.
\eeq
 Here, the $\bar{\theta}_{\rm QCD}$ prediction does not depend on $f_a, f_\f.$ We omit the standard model contribution from the CKM phase. This should be valid when $\bar{\theta}_{\rm QCD}>\O(10^{-17}-10^{-16})$~\cite{Bigi:1991rh,Pospelov:1997uv,Arvanitaki:2014dfa}.\footnote{A detailed estimation of the standard model contribution requires precise calculations of the QCD matrix elements as well as the quark chromo-EDMs and is beyond the scope of this paper. }
 One finds that in our model $\bar{\theta}_{\rm QCD}$ in general is non-vanishing, but it is small if $\min[\L_1^4,\L_2^4]\ll \chi$ with $c_i=\O(1), \theta=\O(1).$
We plot the strong CP phase as a function $\L_1$ and $\L_2$ in Fig.\,\ref{fig:1}. 
$\theta=c_1=1, c_2=0.5, \AND c_3=2$ are fixed. In the blue region, the standard model contribution becomes dominant and our calculation is invalid.

We find that when $\min[\L_1,\L_2]\lesssim {\rm a ~few}\times 10^{-4}\GEV$, $\bar{\theta}_{\rm QCD}$ can be within the experimental bound from the neutron EDM~\cite{Baker:2006ts,Afach:2015sja} $|d_n| \lesssim 3\times 10^{-26}e$\,cm, where the neutron and proton EDMs are related with $\bar{\theta}_{\rm QCD}$ as~\cite{Pospelov:2005pr, Dragos:2019oxn}
\beq 
d_{n}= \(1.52 \pm 0.71\)\times 10^{-16} \bar{\theta}_{\rm QCD} e \,{\rm cm }~\AND~ d_{p}= \(-1.1\pm 1.0\)\times 10^{-16}\bar{\theta}_{\rm QCD} e \,{\rm cm },
\eeq 
respectively.
The bound is represented in the figure by the gray shaded region. Also shown is the 
red band corresponding to $|d_{p}|\approx 10^{-29} e\,$cm, which is the sensitivity reach of a storage ring experiment for measuring the proton EDM~\cite{Anastassopoulos:2015ura}. 
We also show the orange band for $|d_{p}|\approx 10^{-30} e\,$cm which could be achieved in the future.

One can reduce the model to a single-axion one by taking $c_1=0$ and $\L_1^4\to \infty$, i.e. the decoupling limit of $\f.$ 
At the limit, one obtains the potential \beq\laq{singleaxion} V \to -\L_2^4\cos[c_3 a/f_a] -\chi \cos[a/f_a+\theta]+\cdots ~~[{\rm single\text{-}axion~model}] \eeq with certain field and parameter redefinitions. 
Again, there is a CP phase which cannot be shifted away, which induces a non-vanishing strong CP phase. 
The strong CP phase is estimated by taking $c_1=0 \AND \L_1^4\to \infty$ in \Eq{thetaCP} as 
\beq
\bar{\theta}_{\rm QCD}\approx 3\times 10^{-12} c_3^2 \(\frac{\L_{2}}{0.1\MEV}\)^4 \theta~~[{\rm single\text{-}axion~model}].
\eeq
With $\L_2\sim \L_{\rm vis},$ the axion is the QCD axion solving the strong CP problem while inducing a small non-vanishing CP-phase. 

In our scenario, either with single, $\L_1\to \infty, \L_2\sim \L_{\rm vis}$, or two 1\,Hz axions, $\L_1,\L_2\sim \L_{\rm vis}$, the EDMs are predicted around the current bound, and significantly overlap with the future reach.
Therefore, our scenario can solve the strong CP problem, while also giving a testable prediction for small CP-violation in the QCD sector. As we will see, the contribution to $\bar{\theta}_{\rm QCD}$ is also important for testing the scenario using spin-dependent forces.  We mention that further CP-violation may also arise from the MSSM contribution although it depends on SUSY-breaking mediation. In this case, the nucleon EDM contribution from $\bar{\theta}_{\rm QCD}$ can be identified as the minimum value unless there are cancellations between several contributions. 

\subsection{Mass Spectrum and Relic Abundance}
From now on, let us focus $\L_1\sim \L_2 \sim \L_{\rm vis},$ and $c_i=\O(1).$ The single-axion case can be easily obtained by taking the aforementioned decoupling limit of $\f$. 

By diagonalizing the mass matrix, the mass eigenvalues are found to be 
\begin{align}
\laq{mass}
M_H\approx \frac{\sqrt{\chi}}{f_a},~~
M_L\approx\frac{\sqrt{\L_1^4+c_2^2\L_2^4}}{f_\f},
\end{align}
where $M_{H}$ ($M_L$) is the heavier (lighter) axion composed respectively by $a$ ($\f$), i.e. the eigenstates are
\beq
\laq{eigen}
A_H\approx a,~~ A_L \approx \f 
\eeq
with decay constants, respectively, 
\beq
f_H\approx f_a, ~~f_L\approx f_\f. 
\eeq
This formula is valid when $|(c_2 c_1 -c_3)\theta|\lesssim 1.$
Explicitly \beq M_L\sim \(10^{-16}-10^{-15}\)\EV \(\frac{10^{16}{\GEV}}{f_\phi}\)\eeq 
if $\L_1, \L_2 \sim \L_{\rm vis},$ and $c_2=\O(1)$. 
The heavier axion gets its mass mostly from QCD instantons, and is close to the canonical value 
\beq
M_H\approx 6\times 10^{-10}\EV\(\frac{10^{16}\text{ GeV}}{f_L}\).
\eeq
The heavier axion can successfully solve the strong-CP problem, as we have discussed. 

A uniform, non-vanishing initial angle for the axion fields, occurs naturally when the effective symmetry breaking scale is larger than the inflationary Hubble rate. Such a condition must occur for the string/M-theory insipired effective field theory to be under control. Axion fields with non-vanishing initial amplitude start to oscillate around the potential minima when the masses become comparable to the Hubble expansion rate. The oscillation energy contributes to the matter density in the usual vacuum realignment mechanism~\cite{Preskill:1982cy,Abbott:1982af,Dine:1982ah}. 
 The light axion has a dynamical scale $\(\L_1^4+c_2^2\L^4_2\)^{1/4}$, which is temperature independent. Assuming oscillations begin after inflation, and approximating the potential by a quadratic term (See, however, Sec.\,\ref{sec:low_f}), the light axion abundance is 
\beq
\Omega_{L} h^2 \approx 0.1\,
\bigg(\frac{g_{\star,\text{osc}}}{11}\bigg)^{-1/4}
\bigg(\frac{M_{L}}{10^{-15}\,\text{eV}}\bigg)^{1/2} \(\frac{\theta_i^L}{0.2}\)^2\bigg(\frac{f_L}{10^{16}\,\text{GeV}}\bigg)^2. \laq{absa}
\eeq
where $\theta_i^I$ with $I=L,H$ is the initial misalignment angle defined by the initial  amplitude normalized by $f_I$, $g_{\star,\text{osc}}$ is $g_\star$ at the onset of oscillation, and the reduced Hubble rate today, $h$, defined from $H_0=100 h \text{ km s}^{-1}\text{ Mpc}^{-1}$, and we take the value $h\approx 0.68$~\cite{Aghanim:2018eyx}.
The QCD axion has a temperature dependent dynamical scale, and oscillations starts at around the QCD phase transition. The abundance is given by (e.g Ref.~\cite{Ballesteros:2016xej})
\begin{eqnarray}
\laq{QCDaxion}
\Omega^{}_{H} h^2 
\,\approx\, 
0.35 \left(\frac{\theta_i^H}{0.001}\right)^{2}\times
\begin{cases}
\displaystyle 
\left(\frac{f_H}{3\times 10^{17}\,{\rm GeV}}\right)^{1.17} 
& f_H \,\lesssim\, 3 \times 10^{17}\,{\rm GeV} \vspace{3mm}\\
\displaystyle 
\left(
\frac{f_H}{3\times 10^{17}\,{\rm GeV}}\right)^{1.54}
& f_H \,\gtrsim\, 3 \times 10^{17}\,{\rm GeV}
\end{cases}.
\end{eqnarray} 

The total axion abundance, $\Omega_{\rm tot}h^2\equiv \Omega_{ H}h^2 +\Omega_{ L}h^2$, must be less than or equal to than the observed dark matter abundance $\Omega_{\rm DM}h^2= 0.120\pm 0.001$~\cite{Aghanim:2018eyx}, i.e. 
\beq\Omega_{\rm tot}h^2 \leq \Omega_{\rm DM}h^2. \eeq
The total abundance of the two axions with $f_I\sim 10^{16}\GEV$ is too large unless $\theta_i^{L}, \theta_i^H\ll1$, e.g. for the QCD axion abundance alone $\theta_i^H\lesssim 0.01$ is required for $f_H\approx 10^{16}\GEV$. The over abundance problem is solved without fine tuning in Section~\ref{sec:inflation_model}.

\subsection{Axion Couplings}

Both axions can couple to the standard model particles in a general manner. 
The axion couplings to gauge bosons and fermions up to dimension five terms are as follows (at renormalization 
scale below the QCD scale):
\beq
{\cal L}_{\rm int}\supset \frac{\a_e}{ 4\pi}\(c_\g^\f \frac{\f}{f_\f}+c_\g^a \frac{a}{f_a}
\) F\tilde{F}+\( c_n^\f \frac{\partial_\m\f}{f_\phi}+c_n^a \frac{\partial_\m a}{f_a} \)
\ol{\Psi}\gamma^\m\gamma_5\Psi.
\eeq
Here $c_\g^i, c_n^i$ are anomaly coefficients, $F$ and $\tilde{F}$ the photon field strength and its dual, and $\a_e \approx 1/137$  the gauge coupling constant. 
$\P$ represents nucleons, neutrinos and charged leptons, where $c_n^i$ can be different for different fermions. The Lagrangian satisfies the axion shift symmetry (up to total derivatives). 
If the axion couplings to gauge bosons are universal i.e. respect the GUT relation, then $c_\g^a=0.75, c_\g^\f=0$ should be satisfied. However, this condition need not always hold since the GUT breaking mechanism may induce non-universal couplings\, e.g.~Refs.~\cite{Yanagida:1994vq, Izawa:1997he,Kawamura:2000ev, Barbieri:2000vh, Altarelli:2001qj,Hall:2001pg,Ludeling:2012cu}.  
With the general couplings, $A_H\approx a$ and $A_L\approx \f$, both axions couple to photons and fermions. 
Interestingly, $A_L$ can naturally couple to the nucleon/photon without a gluon coupling, i.e. the $A_L$-gluon-gluon anomaly automatically vanishes due to mixing (see \Eqs{intG} and \eq{eigen}).
The couplings provide an opportunity for dark matter detection of $A_L\approx \phi$.

For the photon coupling we consider the sensitivity reaches of the ABRACADABRA~\cite{Kahn:2016aff,Ouellet:2018beu,Ouellet:2019tlz} (see also Ref.~\cite{Salemi:2019xgl} for the latest result) and DANCE~\cite{Nagano:2019rbw, Michimura:2019qxr} experiments  for $c_\g=1$. For ABRACADABRA, seismic noise in broadband is neglected, which also limits the extrapolation to low masses. For the nucleon coupling, we consider the CASPEr-Wind experiment~\cite{Graham:2013gfa,Budker:2013hfa, JacksonKimball:2017elr, Garcon:2017ixh} (see the latest result \cite{Garcon:2019inh})\footnote{See also the constraints from comagnetometers~\cite{Bloch:2019lcy}.}
in optimistic (spin noise only) and conservative (CASPEr-ZULF) cases  for $c^\f_n=1$. In Ref.~\cite{Garcon:2019inh}, the projection is 3 orders of magnitude smaller than Ref.~\cite{Garcon:2017ixh}. The
 $10^3$ increase from the projections in Ref.~\cite{Garcon:2019inh} is obtained assuming hyper polarisation. Furthermore, for low axion masses in all direct searches, stochastic fluctuation of the galactic axion field become important~\cite{Centers:2019dyn} and affect the sensitivity reach by an additional $\mathcal{O}(1)$ factor. 
However, in the 1\,Hz axion window, the coherence time is shorter than a year and the effect may not be very important if a long enough measurement is made. For the CASPEr projections, we consider constant sensitivity extrapolations to low masses. We also mention that the $1\,$Hz axion is right inside the focused mass range of the AION experiment~\cite{Badurina:2019hst}. 

\section{Low-scale inflation as a solution to the overabundance problem}
\label{sec:inflation_model}

We now show, in Section~\ref{sec:basic_tcc}, that the axion over abundance problem can be solved with low scale inflation. In this case, the QCD axion abundance is diluted to nearly zero, and the remaining 1 Hz axion constitutes the entirety of the DM. With GUT scale decay constant, the 1 Hz axion is still out of reach of most experimental searches. We thus present a variant model with lower decay constants which realises the same relative DM abundances in Section~\ref{sec:low_f}. 

One motivation for the low-scale inflation is quantum gravity. 
Recently, by appealing to quantum gravity arguments, it was conjectured \cite{Bedroya:2019snp,Bedroya:2019tba}\footnote{See also Refs.~\cite{Tenkanen:2019wsd,Cai:2019hge,Das:2019hto,Mizuno:2019bxy,Brahma:2019unn,Dhuria:2019oyf,Kamali:2019xnt,Torabian:2019zms,Schmitz:2019uti,Kadota:2019dol,Berera:2019zdd,Goswami:2019ehb,Okada:2019yne,Lin:2019pmj,Li:2019ipk} for subsequent studies on the inflationary cosmology, and argument on the Conjecture~\cite{Saito:2019tkc}.} that the e-folding 
number, $N_e$, during the whole of inflation (larger than the one corresponding to the thermal history of the Universe) is bounded above as
\beq
\laq{TCC}
N_e \lesssim \log[{M_{pl}\over H_{\rm inf}}].
\eeq
This is known as the ``trans-Planckian censorship conjecture'' (TCC): it forbids any mode that ever had wavelength smaller than the Planck scale from being classicalized by inflationary expansion. It protects cosmological observables from sensitivity to trans-Planckian initial conditions. Note that this condition restricts any field value of the potential, e.g. the potential should not have a false vacuum. The TCC predicts inflation scale $H_{\rm inf}\lesssim 1\GEV$~\cite{Bedroya:2019tba}, which is even smaller, $H_{\rm inf}\lesssim 0.01\EV$, for single field slow-roll inflation~\cite{Kadota:2019dol}. We call this inflationary energy scale ``ultra low-scale inflation''.
 In Section~\ref{sec:inflation_model} we build an explicit model for ultra low-scale inflation with $H_{\rm inf}\gtrsim 10^{-15}\,$eV. 
 
 Finally, in Section~\ref{sec:eternal}, we discuss the possibility of low scale eternal inflation.
 
\subsection{Ultra low-scale Inflation and Axion Rolling}


\subsubsection{Diluting the QCD Axion Abundance, and Obtaining 1 Hz Dark Matter}
\label{sec:basic_tcc}

The Hubble parameter for ultra low-scale inflation has Gibbons-Hawking temperature $T_{\rm inf}\approx H_{\rm inf}/2\pi$ smaller than  the QCD scale and thus the QCD axion potential is non-negligible, and the QCD axion rolls, suppressing its abundance dramatically. 

\begin{figure}[!t]
\begin{center}  
     \includegraphics[width=155mm]{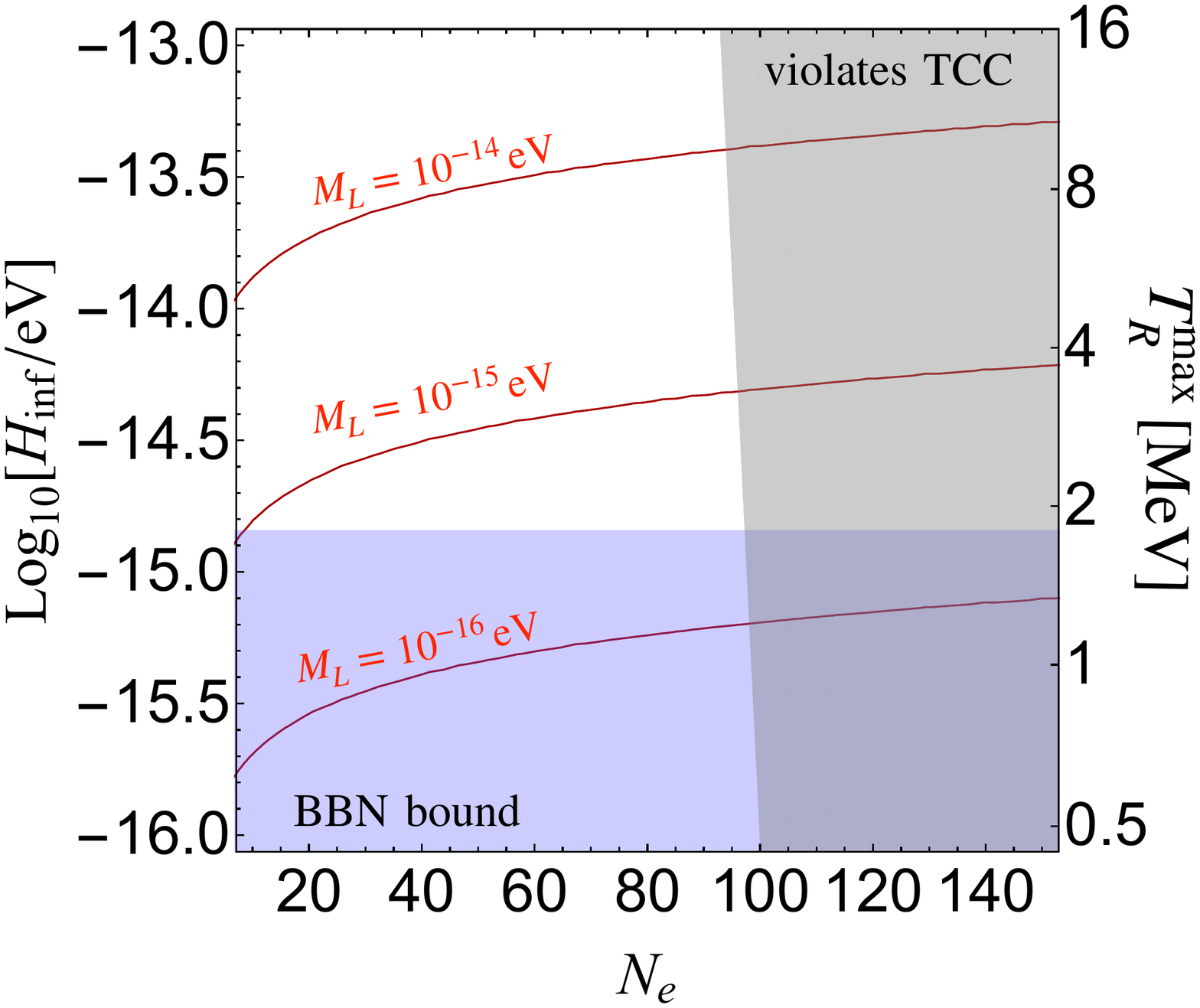}
      \end{center}
\caption{Ultra-low-scale inflation in $N_e-H_{\rm inf}$ plane. Contours have $\Omega_Lh^2=0.12$ for different values of the lightest axion mass $M_L=10^{-14}, 10^{-15}, 10^{-16} \EV$ shown from top to bottom. The maximum reheating temperature is shown on the right hand $y$-axis, and BBN thus gives a lower bound on $H_{\rm inf}$, shown in blue. 
The gray shaded region is inconsistent with the TCC.}\label{fig:2} 
\end{figure}

We are then interested in the lightest axion providing the dark matter, and so we consider $H_{\rm inf}\gtrsim M_L,$ in order not to suppress its abundance too much~\cite{Randall:1994fr}. Thus, during inflation the lighter axion undergoes slow-roll following the equation of motion
\beq
\laq{Eqslow}
\dot{A}_L\approx  \frac{M_L^2}{3H_{\rm inf}} A_L.
\eeq
The solution is 
\beq
\laq{theta1}
\theta_i^L\approx \theta^L_{\rm inf}e^{-\frac{M_L^2}{3H_{\rm inf}^2} N_e},
\eeq
where $\theta_{\rm inf}^I$ is the misalignment angle at the beginning of inflation. 
As a result, $\theta_i^L\ll \theta_{\rm inf}^L=\O(1)$ if ${M_L^2}/{3H_{\rm inf}^2} \gtrsim N^{-1}_e.$ 
In particular if $H_{\rm inf}$ is slightly greater than $M_L$, the lighter axion can remain to explain the dominant dark matter with \Eqs{absa} and \eq{TCC}.
The result after setting $\Omega_{L}h^2= 0.12$ is shown in Fig.\,\ref{fig:2} with $M_L=10^{-14}, 10^{-15}, 10^{-16}\EV$ from top to bottom by taking $\theta_{\rm inf}^L=1$ (i.e. natural misalignment angle
). 
On the right hand $y$-axis we show the corresponding upper limit to the reheating temperature, $T_R^{\rm max} \equiv (g_\star \pi^2/90)^{-1/4}\sqrt{H_{\rm inf}M_{pl}}$. The observational bound is $T_R\gtrsim 2-5\MEV$ from big-bang nucleosynthesis (BBN), depending on the inflaton couplings to the standard model particles~\cite{Kawasaki:1999na,Kawasaki:2000en, Hannestad:2004px, Ichikawa:2006vm, DeBernardis:2008zz, deSalas:2015glj, Hasegawa:2019jsa}. (The blue shaded region denotes the bound adopted from~\cite{Hasegawa:2019jsa} for the leptonic decays of a heavy particle.)
The shaded region is the upper bound to the number of e-folds set by the TCC. 
Given the TCC, the lower bound of $M_L\gtrsim 10^{-16}\EV$ is obtained from the BBN constraint on $T_R^{\rm max}$. (Notice that $N_e\gtrsim N_*$, which is the e-folding after the horizon exit defined later, is needed.)
Consequently, there are parameter regions, where $T_R^{\rm max}\approx \O(1-10)\MEV$, satisfying the BBN bound for a $1$\,Hz axion.

In the viable parameter regions, the heavier ($\approx$QCD) axion energy density is diluted as $\propto a^{-3}\propto \exp[-3H_I t]\propto \exp{(-3N_e)}$ during inflation.  Dilution is caused because $M_H>H_{\rm inf}$, and the heavier axion oscillates during inflation instead of slowly rolling. 
By assuming an instantaneous reheating, $T_R=T_R^{\rm max}$ one obtains the abundance of the heavier axion as 
\beq
\Omega^{\rm TCC}_H h^2=\left.\frac{\rho_{H}}{s}\right|_{T=T_R} \times \frac{s_0}{\rho_c}
\eeq
where $s_0\approx 2900\,{\rm cm^{-3}}, \rho_c\approx 10^5 \GEV h^{-2}\,\text{cm}^{-3}, s=g_{s\star} 2\pi^2 T^3/45$ with $g_{s\star}$ being the relativistic degrees of freedom for the entropy density, and $\rho_H$ is the energy density of the heavier axion. 
Since at the beginning of inflation, $\rho_{H}\lesssim \chi$ should be satisfied, the dilution works as $\rho_{H}|_{T=T_R}\lesssim \chi \exp[-3N_e]. $
Total e-folds satisfy $N_e> N_*=13+\log[T_R^{\rm max}/3\MEV]$, and thus
one obtains the inequality
\beq
\laq{abQCD}
\Omega^{\rm TCC}_H h^2\lesssim \Omega_H^{\rm max} h^2
\eeq
where 
\beq
\Omega_H^{\rm max} h^2\equiv \Omega^{\rm TCC}_H h^2|_{\rho_{H}=\chi \exp[-3N_*]}\sim 10^{-6} \(\frac{3\MEV}{T_R^{\rm max}}\)^6.
\eeq
A more stringent constraint comes from $\rho_H\lesssim H^2_{\rm inf} M^2_{pl}$ at the beginning of inflation i.e. the axion potential does not contribute to the inflation dynamics, or inflation does not start until this condition is satisfied.  
This further suppresses the possible QCD axion abundance by $\sim H_{\rm inf}^2 M_{pl}^2/\chi\ll1$. 
Therefore, the QCD axion abundance is highly suppressed with $T_R^{\rm max}> \O(1)\MEV,$ and $m_a>H_{\rm inf},$ 
and cannot be observed in axion haloscopes~\cite{Graham:2013gfa, Kahn:2016aff, Nagano:2019rbw}~(See also e.g. Refs.~\cite{Asztalos:2009yp, TheMADMAXWorkingGroup:2016hpc,Brubaker:2016ktl, Petrakou:2017epq,Alesini:2017ifp,McAllister:2017lkb, Choi:2017hjy, Marsh:2018dlj, Alesini:2019nzq} for heavier mass range $M_H\gg 10^{-10}\EV$).
Such a QCD axion may, however, be searched for in the ARIADNE force experiment if the decay constant is small enough~\cite{Arvanitaki:2014dfa,Geraci:2017bmq}. In our model, the QCD axion signal in ARIADNE is even enhanced compared to the normal case due to the larger value of $\bar{\theta}_{\rm QCD}$. 

\begin{figure}[!t]
\begin{center}  
     \includegraphics[width=165mm]{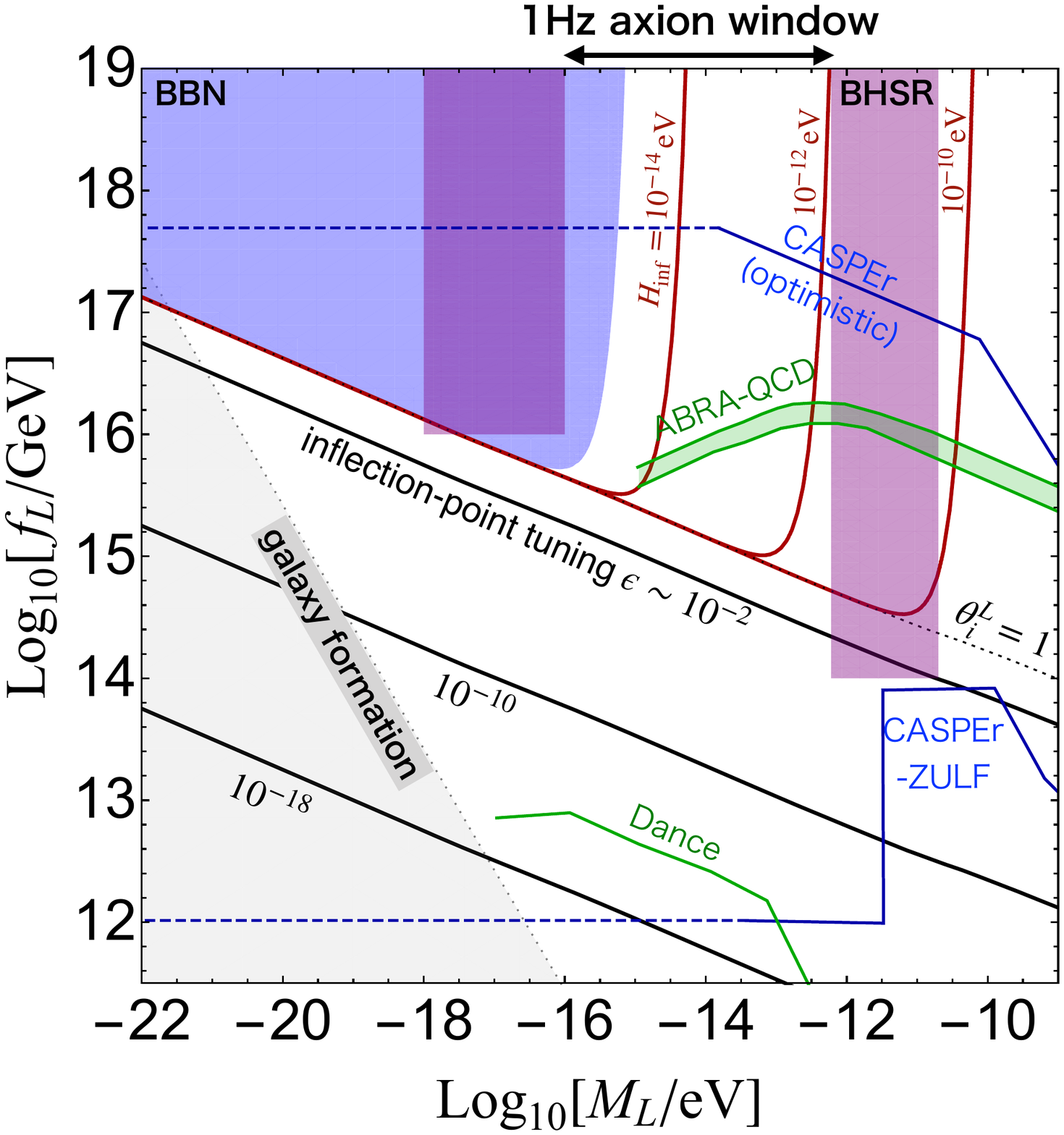}
      \end{center}
\caption{The ($M_L-f_{L}$) 1 Hz axion parameter space fixing with $\Omega_{\rm tot}h^2\approx \Omega_Lh^2= 0.12$ with TCC-satisfying ultra-low-scale inflation. Rolling of the axion fields fixes $\theta_i^I=(H_{\rm inf}/M_{pl})^{M_I^2/3H_{\rm inf}^2}$, and we consider $H_{\rm inf}=10^{-14}, 10^{-12},10^{-10} \EV$ from left to right for the red solid lines. The purple regions and blue region may be excluded by BHSR and BBN constraint, respectively. Below the dotted line, which denotes the standard realignment mechanism, we show the contours for the slope parameter at the inflection point of the effective potential for $\f$ for the correct abundance.   The gray shaded region may be excluded since the oscillation of $\f$ happens with $T\lesssim \KEV.$ 
The lower and upper blues lines are the sensitivity reach of the CASPEr-ZULF experiment with $c^\f_n=1$ taken from~\cite{JacksonKimball:2017elr}, and the optimistic projection taken from Ref.~\cite{Graham:2013gfa} assuming only spin noise, respectively. Dashed lines are our extrapolations. 
Below the upper and lower green band (line), the region can be tested in the ABRACADABRA experiment with $c^\f_\g=1$ adopted from~\cite{slide} and the projected reach of the recently-proposed DANCE experiment~\cite{Nagano:2019rbw, Michimura:2019qxr}, respectively. }
\label{fig:3} 
\end{figure}

In Fig.\,\ref{fig:3}, we show the direct detection prospects with of 1 Hz axion DM, $\Omega_{\rm tot}h^2 \approx  \Omega_Lh^2=0.12$ 
maximising $N_{e}=\log[M_{pl}/H_{\rm inf}]$ (red solid lines for $H_{\rm inf}=10^{-14}, 10^{-12}, 10^{-10}\EV$ from left to right).
 The purple bands are excluded by BHSR, where we impose approximate lower limits 
$f_L<10^{14}\GEV$ and $f_L<10^{16}\GEV$ for the stellar and supermassive regions respectively due to the Bosenova effect~\cite{Yoshino:2012kn} (see Ref.\,\cite{Arvanitaki:2014wva} for the single cosine term case). 
The aforementioned BBN bound is also shown in the blue region. 
Experimental reaches are included as discussed above. 
We conclude that dark matter with $f_L\approx 10^{16}\GEV$ can be mostly tested by ABRACADABRA if $c_\g^\f\gtrsim \O(1)$. 
Furthermore, if the CASPEr experiment can reach the most optimistic sensitivity, then most of the parameter space can also be tested via the nucleon coupling if $c_n^\f=\O(1).$ 

The dotted line represents $\theta^L_i=1$, below which the abundance of the 1 Hz axion is not enough for $\theta^L_i=\O(1)$ and GUT scale decay constant. 
To enhance the abundance dynamically, without tuning the misalignment angle, one possibility is to have a delayed onset of the oscillation. The contours below the dotted line represent the tuning of the potential with a parameter $\epsilon$ to have an inflection point, allowing smaller $f_L$ and thus larger couplings. The details of this mechanism for smaller $f_L$ is as follows.

\subsubsection{Lowering the 1 Hz Decay Constant: Axions at the Inflection Point}
\label{sec:low_f}

Previously we have focused on $f_L\sim 10^{16} \GEV$. 
From Fig.\,\ref{fig:3}, however, we see that a smaller $f_L$ is easier to be tested in the ALP DM experiments due to a stronger interaction. 
From a theoretical point of view, $f_L$ may be smaller or much smaller than $10^{16}\GEV$ due to larger volume of the compactification in the M-theory or a clockwork in the axiverse~\cite{Higaki:2016yqk, Farina:2016tgd}.

From \Eq{absa}, where the potential around the minimum was approximated by a quadratic term, the abundance is suppressed with smaller $f_L$.\footnote{In a single cosine potential, an axion (-like particle) can have an enhanced abundance with anharmonic effect if axion is set at the potential top, $\f/f_\f\approx \pi.$
The enhancement to \Eq{absa} is proportional to $(-\log{|\f/f_\phi-\pi|})^{\O(1)}.$ 
To have a sufficient enhancement we need to tune $|\f/f_\f-\pi|$ exponentially. 
Usually, isocurvature perturbation is significantly enhanced, since the abundance of axion is sensitive to the quantum fluctuation during inflation.  This sets a severe constraint. 
 The both fine-tuning and isocurvature problems can be alleviated with two cosine terms with an almost quartic hilltop potential~\cite{Nakagawa:2020eeg}. 
In our scenario, this may not work since the CP phase, $\theta$, is arbitrary, and the potential may not have an almost quartic hilltop. }
In this part, we will take account of the higher order terms in the potential and show that the abundance of $A_L$ can be enhanced if the axion starts to oscillate from an inflection point.
We also show a mechanism in which the initial condition of $\phi$ is set dynamically in ultra-low-scale inflation satisfying the TCC.

After integrating the QCD axion, $A_H\approx a,$ the potential of $\f$ becomes
\beq
\laq{potL}
V_{L}\approx  -\Lambda_1^4\cos{[\frac{\f}{f_\f}-c_1 \theta]}- \Lambda_2^4\cos{[c_2\frac{\f}
{f_\f} -c_3 \theta ]}.
\eeq
Here we consider $c_2\neq 0,\pm 1$.
At around the potential minimum or maximum, we can approximate the potential by a quadratic term unless $(c_2 c_1 -c_3)\theta$ take specific values such as $0$.

\begin{figure}[!h]
\begin{center}  
   \includegraphics[width=110mm]{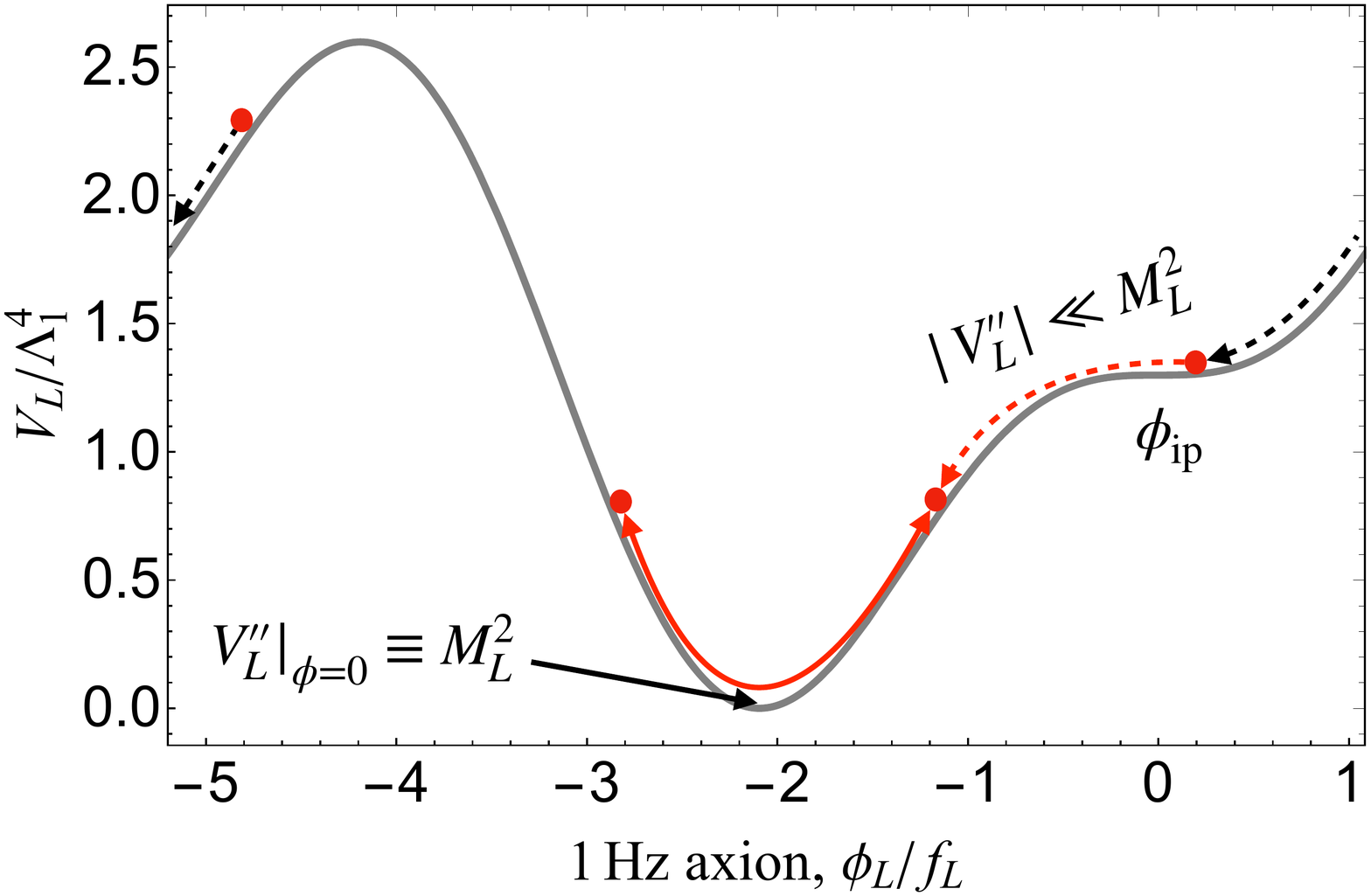}
        \includegraphics[width=110mm]{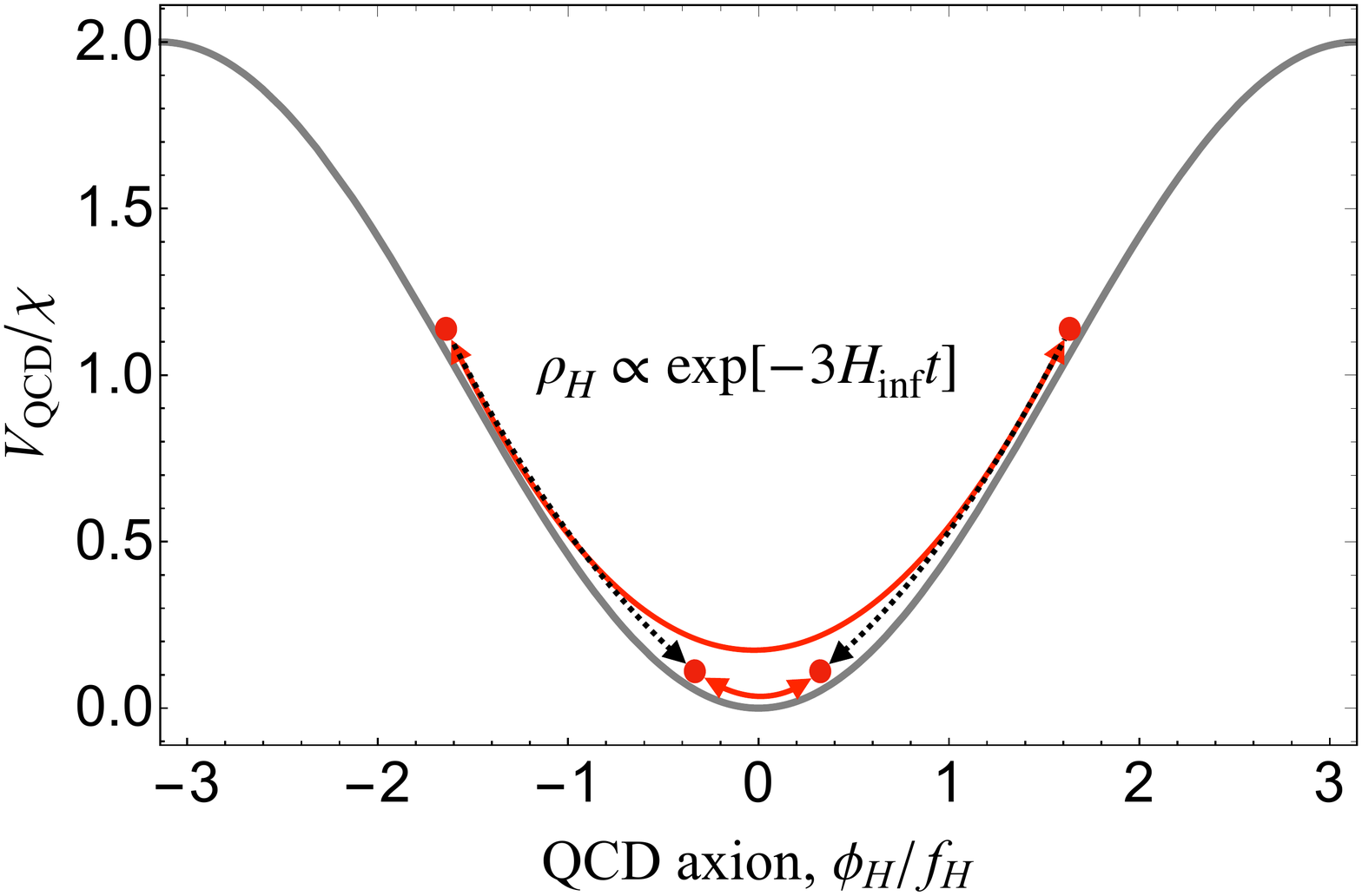}
        \end{center}
\caption{The mechanism for abundant 1\,Hz axion with  $f_L\lesssim 10^{16}\GEV$ [upper panel]. During inflation, the axion is driven to the inflection point, $\phi\approx \phi_{\rm ip}$, with $H_{\rm inf}\sim M_L$ and $N_e< \O(100)$ (Black dashed line). 
After inflation, the onset of oscillation is delayed due to the slow-roll with suppressed slope at around the inflection point (Red dashed line). The oscillation starts at $H^2\sim |V''_L|\ll M_L^2.$ In the lower panel 
we show that the QCD axion starts to oscillate during inflation due to $\L_1^4\ll \chi$. The energy density is diluted due to the exponential expansion [Black dotted line].} 
\label{fig:mech} 
\end{figure}

The inflection point at $\phi=\phi_{\rm ip}$ is defined by $V_L''(\phi_{\rm ip})=0$. The slope of the potential at the inflection point is an important quantity in the discussion. 
In the single cosine case (\eq{potL} with $\L_2\to 0$), this is given as $V_L'|_{\f=\f_{\rm ip}}=\mp \L_1^4/f_\f$ where $\phi_{\rm ip}=\pm \pi f_\f$.
When there are two cosine terms, we can get \beq V_L'|_{\f=\f_{\rm ip}}=\e \L_1^4/f_\f\eeq 
where $\e$ depends on $\L_{2}^4/\L_1^4$ for a given $(c_2 c_1 -c_3)\theta$.
In particular, $|\e|$ can be much smaller than $1$.
For instance, when $c_1=1, c_2=2, c_3=1, \AND \theta=-\pi/2$, 
we can take $\L_2^4= -1/2 (1-\e) \L_1^4$ to obtain 
\beq \laq{inflection}V_{L}\approx V_L^0+\L_1^4\(\sin[\frac{ \f}{f_\f}]- \frac{1-\e}{2} \sin[\frac{2 \f}{f_\f}]\).\eeq
There is an inflection-point around $\f\approx 0$, where $\e\to 0$ with $\L^4_2\to -1/2\L^4_1 $ (see the upper panel of Fig.\ref{fig:mech})

To discuss in detail, let us parameterize the general potential expanded around the inflection point as 
\beq
V_L\approx V_L^0+ \e \L_1^4\frac{\d \f}{f_\f} +a_3{\L_1^4}\(\frac{\d \f}{f_\phi}\)^3+\O\(\frac{\d \f^4}{f^4_\phi}\)
\eeq
where 
\beq
\d \f=(\f-\f_{\rm ip}),
\eeq 
$V_L^0$ a constant term for the vanishing vacuum energy, and $a_3$ is also a function of $\L_{2}^4/\L_1^4, (c_2 c_1 -c_3)\theta$. 
Without loss of generality, we will consider $\f_{\rm ip}> \f_{\rm min}$ and $\e\geq 0$ with $\f_{\rm min}$ being the field value at the potential minimum. 
When $|\e| \ll1$, it is usual that $a_3=\O(1)$ unless $(c_2 c_1 -c_3)\theta$ takes specific values. 
For the example potential \eq{inflection}, $a_3\approx 1/2$.
The linear term dominates the higher order terms when 
\beq
\laq{linear}
|\d \f|\lesssim \ab{\d \f_{\rm crit}}\equiv \sqrt{\ab{\frac{\e}{a_3}}} f_\f.
\eeq

Let us estimate the abundance of $\f$ by assuming that $\f$ is frozen at around the inflection point by the Hubble friction at the end of the inflation, $t=t_{\rm inf}$.
We will show a mechanism to realize this initial condition later. 
When the Hubble friction is important, $\f$ slow-rolls via $\dot{\d \f}\sim -V'_L/(3H)$. We get an equation of
\beq
\laq{slowrollcond}
-\int{d\d\f\frac{1}{V_L'}}\sim \int_{t_{\rm inf}}^{ t_{\rm osc}}{dt\frac{1}{3H} }
\eeq
Here $t_{\rm osc}$ satisfies
\beq
\laq{osc3}
|V_L''| \sim H^2 \eeq
is the time when $\f$ starts to oscillate around the potential minimum. 
Assuming that the oscillation happens at the radiation dominant era, 
we obtain the field value, $\d\f\sim \d\f_{\rm osc}$, at $t=t_{\rm osc}$ from
$ 6 a_3 \frac{\L_1^4}{f_\phi^3} \ab{\d \f_{\rm osc}} \sim \frac{1}{4t^2_{\rm osc}}.$

The l.h.s  and r.h.s of \Eq{slowrollcond} are estimated
$
 \min{[|\d \f_{\rm osc}|, |\d \f_{\rm crit}|]}\frac{f_\phi}{\e \L_1^4},
$
and 
$ \frac{2}{3} t_{\rm osc}^2,$ respectively. 
From \Eqs{slowrollcond} and \eq{osc3}, we obtain
\beq
t_{\rm osc}\sim \frac{1}{\e^{1/4} M_L},
\eeq
where we have ignored the dependence of several $\O(1)$ coefficients, and replaced
$
 \O\(\frac{\L_1^4}{f_\phi^2}\)
$
with the axion mass $M_L^2$.

Compared with the quadratic potential case, $t_{\rm osc}\sim 1/M_L,$ the onset of oscillation is delayed by $1/\e^{1/4}$ for a given $M_L.$ 
This means that the oscillation happens when the entropy density is smaller than the quadratic case by a factor of $\e^{3/8}$. 
As a result the abundance is enhanced as
\beq
\Omega_{L} h^2 \sim 0.3 \(\frac{10^{-4}}{\e}\)^{3/8}\,
\bigg(\frac{g_{\star,\text{osc}}}{11}\bigg)^{-1/4}
\bigg(\frac{M_{L}}{10^{-15}\,\text{eV}}\bigg)^{1/2} \bigg(\frac{f_L}{10^{14}\,\text{GeV}}\bigg)^2, \laq{ab3}
\eeq
where we have assumed $\f_{\rm ip}/f_L=\O(1).$ 
The contours of $\e$ for the correct abundance is shown in Fig.\,\ref{fig:3}.
Note that the onset of oscillation should be much before the structure formation~\cite{Sarkar:2014bca}. 
Much below the gray shaded region may be excluded from galaxy formation since the oscillation starts at  $T\ll \KEV$ (around the line could have implications on small scale structure problems~\cite{Agarwal:2014qca}). 
 \\

Next we show as one example mechanism that
when $|\e|\ll 1$, $\f$ can be driven to the inflection point, $\f \approx \f_{\rm ip}$, during inflation with $H_{\rm inf}\sim M_L$ and with $N_e \lesssim \O(100)$. 
  Suppose that at the beginning of the inflation $\f>\f_{\rm ip}$ and the curvature is of the order $\L_1^4/f_\phi^2=\O(M_L^2)$. 
During inflation, $\f$ slow-rolls and approaches to $\f_{\rm ip}$  with $H_{\rm inf}\gg M_L$. If $H_{\rm inf}\sim M_L$ in a short period, $\f$ reaches $\f_{\rm ip}.$
At around $\f\approx \f_{\rm ip}$ both the curvature and slope are suppressed, and $\dot{\f}$ becomes slower.  
One can estimate the field excursion around the inflection point as a function of the period of e-folding number, $\D N_e$, as
\beq
\ab{\frac{\D\f}{f_\f}} \sim \frac{|\e|}{3} \D N_{e} \frac{\L_1^4}{f_\f^2 H_{\rm inf}^2}.
\eeq
For $\D N_e=\O(10)$, which may be  restricted from the TCC, and $|\e|\ll 1$, $\f$ cannot pass through the whole region of \eq{linear}.\footnote{
Another possibility is that $\f$ oscillates at the beginning of inflation with a sufficiently large amplitude. 
During the inflation the amplitude decreases. 
When $\f$ happens to be the value in the inflection point regime, $|V''_L| \lesssim H_{\rm inf}^2$, with a small enough $|\dot{\f}|$, 
the oscillation stops. The initial condition is set. }
As a result, during the inflation $\f$ is driven to the inflection point. 
 
 The whole mechanism is summarized in Fig.\,\ref{fig:mech}. We also show the oscillation of the QCD axion during inflation in the lower panel. Again since $M_H\gg H_{\rm inf}$ the QCD axion starts to oscillate during inflation and the abundance is diluted. 
 
 Finally, let us have four comments on the scenario. First, the isocurvature perturbation is highly suppressed. This is because the axion abundance, \eq{ab3}, is not so sensitive to the initial amplitude of $\f$. 
  In this case, the quantum fluctuation of $\f$ during the inflation would not lead to significant density perturbation of the axion. 
 Second, in the context of TCC, the whole potential of $\f$ should not have a false vacuum, which would lead to an ``old" eternal inflation. 
 The previous discussion can be consistent with the absence of a false vacuum, e.g. as in Fig.\,\ref{fig:mech}. 
Also, the TCC on the inflection-point inflation driven by $\f$ around $\f_{\rm ip}$ is easily satisfied.

Third, if $c_2$ is not much larger than $\O(1)$, $\L_2\sim \L_1$ must be satisfied to have an inflection point (otherwise it would be effectively a single cosine potential). In this case, we need to make sure $\L_1< \O(1\MEV)$ to evade the EDM bound in Fig.\,\ref{fig:1}. 
This sets an upper bound for $M_L$ with $|(c_2 c_1 -c_3)\theta|= \O(1)$ as 
\beq
M_L \lesssim 10^{-9}\EV \(\frac{10^{12}\GEV}{f_L}\).
\eeq
This scenario can be tested if 
\beq
M_L \gtrsim 10^{-13}\EV \(\frac{10^{12}\GEV}{f_L}\).
\eeq
Four, although our discussion is motivated by the TCC, the mechanism should also work in thermal inflation scenario~\cite{Yamamoto:1985rd,Lyth:1995ka}, which also has a short period of the exponential expansion of the Universe.

\subsection{A Model for Ultra-low-scale inflation: higher-order-inflection-point inflation}
\label{sec:inflation_model}

Recently, ALP inflation with $H_{\rm inf} \gtrsim \O(10^{-6})\EV$ was studied with two cosine terms composing the inflaton potential~\cite{Daido:2017wwb, Daido:2017tbr, Takahashi:2019qmh} (see also Refs.~\cite{Czerny:2014wza, Czerny:2014xja} for multi-natural inflation).  
In particular in Ref\,\cite{Takahashi:2019qmh}, it was discussed that in the regime of inflection-point inflation with a potential similar to \Eq{inflection} where the inflaton potential is dominated by a cubic term, 
the inflationary period has an upper bound. This is a good property to satisfy the TCC. However, the inflation scale may be still too high for our purpose. 

To further lower the inflation scale, one may encounter two problems in general. 
The first is the constraint from the cosmic microwave background (CMB) on the primordial power spectrum. 
The slow-roll period of inflation from the horizon exit of the CMB scales is $N_*\lesssim 20$ for $H_{\rm inf}\lesssim 10^{-6}\EV$. 
The curvature of the inflaton potential has to change fast enough to terminate the slow-roll. 
The derivatives of the curvature at the horizon exit induce running of the scalar spectral index, which is constrained from the CMB data. 
Another potential problem is reheating. 
The inflaton mass may be too small to reheat the Universe, or couplings to the standard model particles for the reheating may be constrained from various experiments.  
We will show that inflection-point inflation driven by a higher-order term than the cubic one can have a very low inflation scale simultaneously consistent with the CMB observations and the TCC bound, and thus realise the desired axion relic abundances in our model. 

\begin{figure}[!t]
\begin{center}  
   \includegraphics[width=145mm]{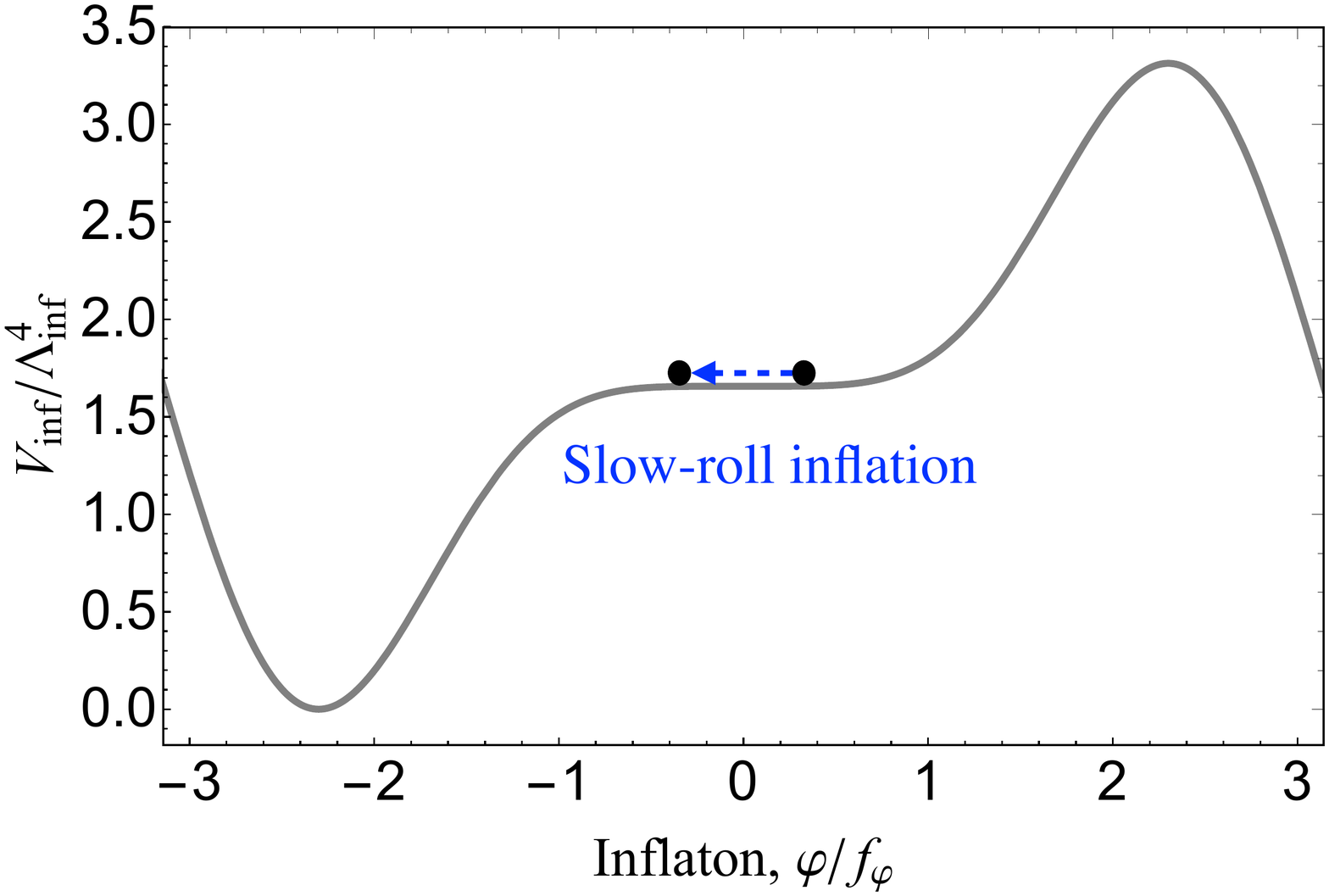}
      \end{center}
\caption{Ultra low-scale inflaton potential, satisfying the TCC. Inflation takes place at $\varphi\approx 0.$ The direction of slow-roll is shown by the blue dashed line.} 
\label{fig:vinf} 
\end{figure}

For concreteness, let us consider inflation driven by a dimension five term.
To control the whole field space let us assume the inflaton is a third ALP, $\varphi$, the potential for which has a discrete shift symmetry with $\varphi\to \varphi +2\pi f_\varphi$. Here $f_\varphi$ is the instanton decay constant, i.e. the potential is periodic on $\varphi$. Thanks to the symmetry, any Coleman-Weinberg corrections are not allowed and thus the slow-roll conditions will be guaranteed at quantum level. 
For instance, we can write down a potential made up by three sine functions.
\beq
\laq{potinf}
V_{\rm inf} = \L_{\rm inf}^4\((1+\d) \sin\(\frac{\varphi}{f_\varphi}\)- \frac{4}{5}\sin\(2\frac{\varphi}{f_\varphi}\)+\frac{1}{5}\sin\(3\frac{\varphi}{f_\varphi}\)\)+V_0
\eeq
Here $V_0\approx 20\sqrt{5} \L_{\rm inf}^4/27$ is for the vanishingly small cosmological constant, $\d \ll 1$ is a small parameter needed to explain the CMB data and also will be important to satisfy the TCC bound. The potential is shown in Fig.\,\ref{fig:vinf} at the limit $\d\to 0$. One can find the only position that may lead to inflation is around $\varphi\approx 0. $ 
  At around $\varphi=0$ one obtains
 \beq
 V_{\rm inf}\approx V_0+\frac{ \L_{\rm inf}^4}{5} \(\frac{\varphi}{f_\varphi}\)^5+\d  \L_{\rm inf}^4\frac{\varphi}{f_\varphi} +\cdots,
 \eeq	
where we have written down the leading terms of $\varphi$ proportional to $\d^0 \AND \d^1.$ 
To have this kind of specific form, we have implicitly tuned two relative hight and two relative phases of sine terms, to have small enough $\varphi^{1,2,3,4}$ terms. 
Although we set the potential shape by hand, we can get a similar form from extra-natural inflation with properly charged matter particles~\cite{Croon:2014dma}. 

At the vanishing limit of $\d$, a $\varphi^5$ term drives the slow-roll inflation. 
Since the curvature term increases, as $|d V_{\rm inf}''/d\varphi|\propto |\varphi|$, one can have smaller $|V'''_{\rm inf}|$ at the horizon exit of the CMB scales, $|\varphi|\approx |\varphi_*|$, than the typical value during the slow-roll to increase $|V''_{\rm inf}|$ to $|V''_{\rm inf}|\sim H_{\rm inf}^2$ at the end of inflation. (Recall that $|\varphi_*|$ is much smaller than the typical value of $\ab{\varphi}$ around the end of inflation.) 
At the limit $\d\to 0$, one can estimate the slow-roll parameters at the horizon exit: $\varepsilon \equiv \frac{M_{pl}^2}{2} \left(\frac{V_{\rm inf}'}{V_{\rm inf}}\right)^2\approx \frac{729 M_{pl}^2 \varphi_*^8}{4000 f_\varphi^{10}},
\eta  \equiv M_{pl}^2 \frac{V''_{\rm inf}}{V_{\rm inf}} \approx \frac{27M_{pl}^2\varphi_*^3}{5\sqrt{5}f_\varphi^5}.$
The predicted power spectrum of the curvature perturbation is given by $P_R(k_*)\approx \frac{5 \sqrt{5}  \L_{\rm inf}^4}{162 \pi^2\e M_{pl}^4}$. 
From the slow-roll equation of motion $3H \dot{\varphi}\approx -V_{\rm inf}'$,  $\varphi_*$ is related to the e-folding number after the horizon exit, $N_*$:
\beq
N_*\approx -\frac{20 \sqrt{5} f_\varphi^5}{81{M_{pl}^2 \varphi_*^3}}.
\eeq
From the CMB normalization condition, $P_R\approx 2.1\times 10^{-9}$, the typical inflation scale $ \L_{\rm inf}$ is predicted as
\beq
\laq{CMBnom}
 \L_{\rm inf} \approx 6.2\MEV  \(\frac{{f_\varphi}}{1\MEV}\)^{5/6} \(\frac{15}{N_*}\)^{2/3}.
\eeq
The slow-roll parameters are derived as
\begin{align}
\varepsilon&\approx  1.4\times10^{-76} \(\frac{{f_\varphi}}{1\MEV}\)^{10/3} \(\frac{15}{N_*}\)^{8/3}\\
\eta&\approx -\frac{4}{3 N_*}.
\end{align}
The tensor/scalar ratio, $r\simeq16 \varepsilon$ is extremely small which is consistent with the CMB data.
The spectral index of the scalar perturbation is given as 
\beq
\laq{ns}
n_s\approx 1-6\varepsilon+2\eta \approx 1-\frac{8}{3N_*}.
\eeq
One finds that this would be in tension with the observed value~\cite{Akrami:2018odb} $n_s^{\rm CMB}= 0.9649 \pm 0.0044$ ({\it Planck}  TT,TE,EE+lowE) since \beq N_*\sim 12 +\log{\(\frac{ \L_{\rm inf}}{1\MEV}\)},\eeq
set by the thermal history by assuming instantaneous reheating. 
However, a better fit of the CMB data can be obtained with a non-vanishing but tiny $\d$. 
This is because  the tiny linear term slightly shifts $\varphi_*$ for a given $N_*$ to enhance $n_s$~\cite{Takahashi:2013cxa}. 
The tiny $\d$ will be also important to satisfy the TCC bound. 

Let us perform  numerical analysis on the inflation dynamics following Refs\,\cite{Daido:2017tbr, Takahashi:2019qmh} up to the third order of the slow-roll expansion. 
The input parameters are \beq \laq{input} f_{\varphi}=0.3\MEV\AND \d=0.0365 \times \(f_\varphi/M_{pl}\)^{8/3}.\eeq The result is as follows 
\beq
H_{\rm inf}\approx 5\times 10^{-15}\EV, n_s\approx 0.964, \a_s\approx -0.0124, \AND \b_s\approx -0.00358,
\eeq
where $ ( \a_s, \b_s)$ is (the running of $n_s$, the running of $\a_s$). 
This is consistent with the result of {\it Planck} TT,TE,EE+lowE~\cite{Akrami:2018odb}.

Next let us consider reheating. 
We need the reheating temperature $T_R \gtrsim 1\MEV$ to have a successful BBN given the correct baryon asymmetry. 
To have reheating let us couple the inflaton to a neutrino pair as 
\beq
i\frac{ c_\nu m_\nu}{f_\varphi} \varphi\bar{\nu^c}\g_5\nu,
\eeq
One can identify $\varphi$ as a Majoron corresponding to the lepton number $-2/c_\nu.$ 
 The decay rate is 
\beq
\G_\varphi\approx \frac{m_\varphi}{4\pi} \frac{c_\nu^2 m_\nu^2}{f_\varphi^2},
\eeq
where \beq 
m_\varphi\approx 2.2 \frac{\L_{\rm inf}^2}{f_\varphi} \approx 85\MEV \(\frac{15}{N_*}\)^{4/3} \(\frac{f_\varphi}{1\MEV}\)^{2/3}\eeq
 and $m_\nu$ is the neutrino mass. We have used \Eq{CMBnom} to estimate the mass. 
 Then the decay rate is 
\beq
\G_\varphi\approx 1.7\times 10^{-8}\EV \(\frac{1\MEV}{f_\varphi/c_\nu}\)^{4/3} \(\frac{15}{N_*}\)^{4/3} \(\frac{m_\nu}{0.05\EV}\)^2.
\eeq
The requirement of $T_R\gtrsim 1\MEV$ turns out to be 
\beq
f_{\varphi} \lesssim 200\GEV \(\frac{c_\nu m_\nu}{0.05\EV}\)^{3/2} \(\frac{15}{N_*}\).
\eeq
where we approximate $T_R= (g_\star \pi^2/90)^{-1/4}\sqrt{\G_{\varphi}M_{pl}/3}.$
Since the  ``Yukawa" coupling to the neutrino $\sim  c_\nu m_\nu/f_\varphi\approx 5\times 10^{-8}c_\nu (m_\nu/0.05\EV) (1\MEV/f_\varphi)$ is tiny, the model 
can be consistent with various bounds for the Majoron coupling. 
The most severe one may be from the SN1987A~\cite{Heurtier:2016iac}. A tiny parameter region may be in tension with the observations, and the model may have implications on 
future supernova observations such as in IceCube.

\begin{figure}[!t]
\begin{center}  
   \includegraphics[width=105mm]{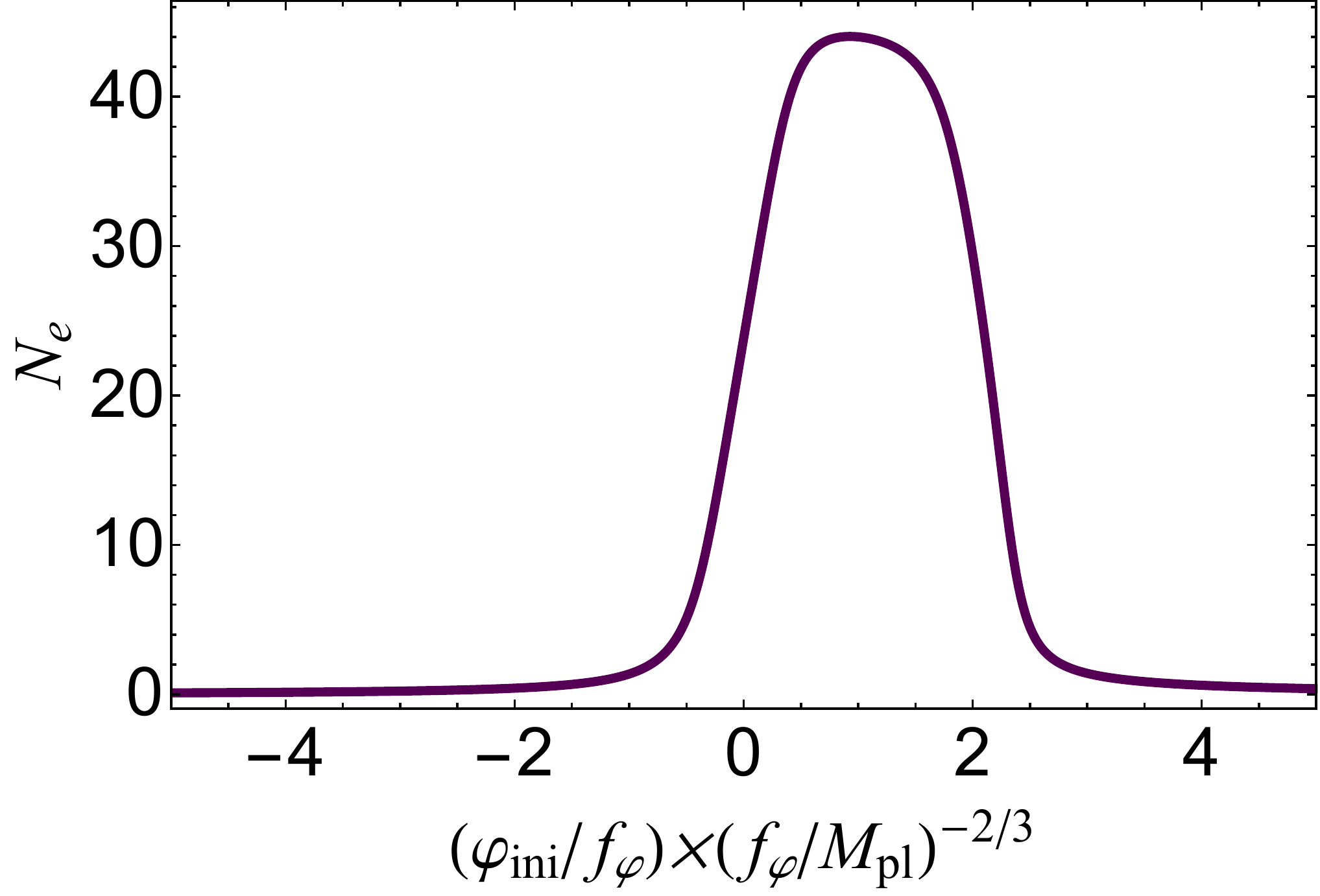}
      \end{center}
\caption{Total e-folding number obtained by varying $\varphi_{\rm ini}$. Here $f_\varphi=0.3\MEV, \d=0.0365 \times \(f_\varphi/M_{pl}\)^{8/3}$ which results $H_{\rm inf}\approx 5\times 10^{-15}\EV. $ } 
\label{fig:efold} 
\end{figure}

Now let us discuss whether the model can be consistent with the TCC. 
The total e-folding number by varying the initial field value, $\varphi=\varphi_{\rm ini}$, around $\varphi=0$ is shown in Fig.\,\ref{fig:efold} for the parameter set \eq{input}.  Since the maximum of the e-folds for this potential (which should not be confused with $N_*$) is around $40-50$, the TCC, which restricts $N_e\lesssim 100$ for $H_{\rm inf}\sim 10^{-15}\EV$, is satisfied. 
We notice that the TCC can be satisfied due to the not extremely small $\d$ to enhance the spectral index. 
If $\d$ were too small, the slope around the inflection point would be too small and leads to too long inflation. 
This inflation model is consistent with the ultra-low-scale inflation needed for opening the $1\,$Hz axion window and consistent with the TCC bound.

We also mention that with $T_R$ smaller than the EW scale, low-scale baryogenesis from inflaton decays is possible~\cite{Dimopoulos:1987rk, Babu:2006xc, Grojean:2018fus, Pierce:2019ozl, Asaka:2019ocw } (one can also transfer the baryon asymmetry to baryonic dark sector from the inflaton decays e.g. Refs\,\cite{Barr:1990ca, Kaplan:1991ah, Kitano:2004sv, Kaplan:2009ag, Eijima:2019hey}).

\subsection{Low scale eternal inflation}
\label{sec:eternal}

In general, there are fine-tunings on the inflaton field over certain spacetime volumes to start inflation. 
Such a tuning may be explained by eternal inflation~\cite{Linde:1982ur,Steinhardt:1982kg,Vilenkin:1983xq}, which predicts extremely large e-folding number, $N_e$. Eternal inflation is obviously in contradiction with the TCC and in the following we abandon it. In this case, we can go further into the shaded region of Fig.\,\ref{fig:2}, where higher $H_{\rm inf}$ can be allowed with larger $N_{e}$. 

In fact, there is an asymptotically allowed value once we take the $N_e\rightarrow\infty$ eternal inflation limit. This is because if we take  $N_e$ extremely large with a given $H_{\rm inf}\gg M_L$, the abundance of $A_L$ saturates due to quantum diffusion. 
It was pointed out that if the inflation scale is sufficiently low and the period is long enough,  the initial misalignment angle of the QCD axion~\cite{Graham:2018jyp,Guth:2018hsa} and string axions~\cite{Ho:2019ayl} reach equilibrium between the classical motion and 
quantum diffusion for the time scale $N_e\gtrsim {H_{\rm inf}^2}/{M_I^2}$. In this case it is the Bunch-Davies (BD) distribution that determines the misalignment angle, which is independent from $\theta_{\rm inf}^I.$
The typical misalignment angles in this ``natural" region are given by 
\beq
\laq{BDi}
\theta_i^{I} \approx \sqrt{\frac{3}{8\pi^2}}\frac{H_{\rm inf}^2}{M_{I}f_I} ~~(N_e\gtrsim \frac{H_{\rm inf}^2}{M_I^2})
\eeq
where we take the variance of the BD distribution. The probability that $\theta_i^I$ is much greater than the variance is exponentially suppressed since the BD distribution is a normal distribution. 
In this model, and taking account of the BD distribution, 
$H_{\rm inf}=\KEV-\MEV$ is possible to have $\Omega_{L}h^2= 0.12$ with $f_L\gtrsim 10^{16}\GEV$. 
We notice again that the dominant dark matter is $A_L\approx \phi$, and the QCD axion $A_H\approx a$ has negligible abundance with $H_{\rm inf}\lesssim 1 \MEV$\,\cite{Graham:2018jyp,Guth:2018hsa}. 
The predictions for direct detection are thus the same as the short inflation case. 

In this model variant, the required $H_{\rm inf}$ can be much larger than the TCC scenario, and thus higher reheating temperature can be obtained. Model-buildings of inflation and baryogenesis~\cite{Pilaftsis:1997dr,Buchmuller:1997yu,Akhmedov:1998qx,Asaka:2005pn,Lazarides:1991wu,Asaka:1999yd, Hamaguchi:2001gw,Hamada:2018epb,Eijima:2019hey} is much easier than in the TCC model. For the inflation, 
one can also consider the previous inflation model.

It is also possible to have the dominant dark matter of $A_L$ with $f_L\ll 10^{16}\GEV$. 
This is the case by setting the axion around the hilltop of the potential due to the mixing between the inflaton and the axion \cite{Daido:2017wwb,Takahashi:2019pqf} (see also a CP symmetric MSSM scenario to set the QCD axion on the hilltop~\cite{Co:2018mho}.), which shifts the potential. 
If the axion is set to the quadratic hilltop, the scenario again suffers from the isocurvature problem with a too small $f_L$. Similarly the axion can be set at around the inflection point. For details of the $\pi$nflation see Refs.\,\cite{Takahashi:2019pqf,Nakagawa:2020eeg}.

Finally, let us connect the two parameter regions of $N_e$. 
Given a $N_e$ in the range of $\log[\frac{M_{pl}}{H_{\rm inf}}] \lesssim  N_{e} $,   \beq M_L \lesssim H_{\rm inf}\lesssim \O(\MEV) \laq{Hinftot}\eeq to explain the abundance of dark matter in general. 
The last inequality becomes an equality if $N_e\gtrsim H_{\rm inf}^2/M_L^2$.

\section{Discussion and conclusions}
\label{sec:discussion}

We have studied a two axion model inspired by M-theory with two would-be $1\, $Hz axions. 
By turning on the gluon coupling, one of the axions becomes more massive and becomes 
the QCD axion, while the other remains light. Introducing more axions in the 1 Hz window does not affect our conclusions. 

With the expected GUT scale decay constants, this model has a relic abundance problem, and we introduced two models to suppress it. In the ultra-low scale scenario, the abundance of the heavier QCD axion is diluted 
with $\exp{(-3N_e)}$, since it oscillates during inflation. Thus the lightest axion tends to contribute dominantly to 
the dark matter and the heavier ones almost do not contribute (an exception is when the axions, including the lightest one, have almost degenerate masses). 
We presented an explicit model for ultra low-scale inflection-point inflation driven by a third ALP.

On the other hand, for the low-scale eternal inflation scenario, from \Eqs{BDi} and \eq{absa}, the lightest axion also 
contributes most to the total abundance in the Bunch-Davies distribution. 
Since the dependence of the abundance on the axion mass is 
not exponential, the inclusion of additional axions may decrease the required  
$H_{\rm inf}$ slightly. However, this does not pose serious problems since 
$H_{\rm inf}$ is not very low and reheating does not present a problem in this model. 

In both inflation models, the 1 Hz axion DM can be searched for with ABRACADABRA, and in the most optimistic scenarios also by CASPEr. The photon coupling may also be tested by future measurements of CMB spectral distortions, if primordial magnetic fields on Mpc scales are nG or stronger~\cite{Tashiro:2013yea}. Detection prospects improve if the axion decay constant is lowered. In this case there is an under abundance problem for the 1 Hz axion, which we showed can be solved with a mild fine tuning of the axion potential introducing an inflection point.

Notice that once either axion coupling as well as the mass of $A_L$ is measured, the existence of the QCD axion is anticipated. This is because if there were no other axion composing the QCD axion, $A_L$ would acquire $\O(10^{-10})\EV$ mass and become the QCD axion unless the coupling to gluons is highly suppressed. In the case of $A_L$ discovery in the 1 Hz axion window (including the region with $f_L \ll 10^{16}\GEV$), one may make an effort to measure the QCD-axion mediated force in the aforementioned ARIADNE axion experiment, or measure the nucleon EDM, to further confirm our model. 

Let us mention the alleviation of other cosmological problems in the UV model when a low inflationary scale is adopted. 
The moduli, if stabilized by SUSY breaking, have masses around  $m_{3/2}$. If such moduli are displaced from the potential minima during inflation, they later dominate the Universe via coherent oscillations and cause the notorious cosmological moduli problem~\cite{Coughlan:1983ci,deCarlos:1993wie}. In our scenario, obviously $H_{\rm inf}\ll m_{3/2}$ is satisfied. The moduli oscillations start during inflation and the abundances are exponentially diluted.
Therefore the moduli problem for the mass $\sim m_{3/2}$ is solved. On the other hand there may be a ``moduli'' problem induced by lighter axions if some axions happen to have masses in the range $10^{-22}-10^{-18}\EV$ and the decay constants $\gtrsim 10^{17}\GEV$. The abundance of these lighter axions is overproduced with the values of $H_{\rm inf}$ required to explain the 1\,Hz axion dark matter abundance (see Ref.\,\cite{Ho:2019ayl}). 
However, such axions may be absent if the mass distributions take particular shapes or there is a small total number of axions~\cite{Stott:2018opm}. 
Another implicit cosmological problem is the gravitino problem. For $\O(1)\TEV\lesssim m_{3/2}\lesssim 50\TEV$, gravitino decays may spoil BBN. 
Thus the gravitino abundance, which is produced most efficiently at high temperatures, should be small enough at its decay. 
This sets an upper bound on the reheating temperature $T_R\lesssim 10^{5-9}\GEV$~\cite{Kawasaki:2017bqm}. 
This is easily (absolutely) consistent with our (ultra) low-scale inflation scenario. 
These facts imply that the moduli and SUSY breaking scales can be lower than the traditional 10-100 TeV bound from the moduli and gravitino problems (there are of course collider bounds setting SUSY scale higher than $\O(1)\TEV$). 

In conclusion, we have opened the window of 1\,Hz axions which it has been suggested are a natural prediction of the M-theory axiverse~\cite{Acharya:2010zx,Stott:2017hvl}. We showed that a would-be 
1\,Hz axion can be the QCD axion and solve the strong CP problem while at the same time inducing a testable non-vanishing strong CP phase. The abundance of the axions can be consistent with 
the observed one, while maintaining GUT-scale decay constants (or much smaller) with sufficiently low-scale inflation. Since the $1$\,Hz axion 
dark matter is dominant, the scenario predicts non-observation of QCD axion dark matter but many phenomenological observables in the $1$\,Hz window from the lighter field.

\section*{Acknowledgments}
We acknowledge useful discussions with Dmitry Budker, Derek Jackson Kimball, Yonatan Kahn, Yannis Semertzidis, Yunchang Shin and Fuminobu Takahashi. 
We are also grateful to Fumi for reading our manuscript. 
WY thanks Institut fur Astrophysik at Georg-August-Universit\"{a}t for kind hospitality when this work was initiated. 
DJEM is supported by the Alexander von Humboldt Foundation and the German Federal Ministry of Education and Research.
WY is supported by JSPS KAKENHI Grant Number 15K21733 and 16H06490,  and NRF Strategic Research Program NRF-2017R1E1A1A01072736.

\end{document}